\documentclass[aps,preprint,showpacs,floatfix]{revtex4}
\pdfoutput=1
\usepackage{graphicx,bm,amsmath,dcolumn}
\RequirePackage{color}

\definecolor{MyDarkGreen}{rgb}{0.02,0.60,0.06}

\begin{document}

\title {Logarithmic negativity 
  of the 1D antiferromagnetic spin-$1$ Heisenberg model with single-ion anisotropy}
	
	\author{Vl. V.\ Papoyan}
	\email{vpap@theor.jinr.ru}
	\affiliation{Bogoliubov Laboratory of Theoretical Physics, Joint Institute for Nuclear Research, 141980 Dubna, Russian Federation}
	\affiliation{Dubna State University, Dubna, Russian Federation}
    \author{G.\ Gori}
	\email{gori@thphys.uni-heidelberg.de}
    \affiliation{Institut f\"ur Theoretische Physik, Universit\"at Heidelberg, 69120 Heidelberg, Germany}
    \affiliation{CNR-IOM DEMOCRITOS Simulation Center, Via Bonomea 265, I-34136 Trieste, Italy}
	\author{V. V.\ Papoyan (Jr.)}
	\email{papoyan8@gmail.com}
    \affiliation{Meshcheryakov Laboratory of Information Technologies, Joint Institute for Nuclear Research, 141980 Dubna, Russian Federation}
	\affiliation{Dubna State University, Dubna, Russian Federation}
	\author{A.\ Trombettoni}
	\email{atrombettoni@units.it}
    \affiliation{Department of Physics, University of Trieste, Strada Costiera 11, I-34151 Trieste, Italy}
	\affiliation{CNR-IOM DEMOCRITOS Simulation Center, I-34136 Trieste, Italy}
    \author{N.\ Ananikian}
	\email{ananik@yerphi.am}
    \affiliation{National Science Laboratory, Alikhanian Br. 2, 0036 Yerevan, Armenia}
	\affiliation{CANDLE Synchrotron Research Institute, Acharyan 31, 0040 Yerevan, Armenia}

	
	\begin{abstract}
          We study the 1D antiferromagnetic spin-$1$ Heisenberg XXX model
          with external magnetic field $B$ and single-ion anisotropy $D$
          on finite chains. 
          We determine the nearest and non-nearest neighbor
          logarithmic entanglement $LN$. Our main result is the
disappearance of $LN$ both for nearest and non-nearest neighbor
(next-nearest and next-next-nearest) sites at zero temperature and
for low-temperature states.
Such disappearance occurs at a critical value of $B$ %
and $D$. 
The resulting
            phase diagram for the behaviour of LN is discussed in the $B - D$ plane,
           including a separating line -- ending in a triple point --
          where the energy density is independent on the size. Finally, results for $LN$ at finite temperature
          as a function of $B$ and $D$ are presented and commented.
	\end{abstract}

	\maketitle
	
\section{Introduction}
\label{Introduction}

Quantum fluctuations and competing interactions in many-body quantum systems 
play a crucial role because they contain important information on various physical observables,
leading to unusual and exciting physical properties connected to the occurrence of strong correlations \cite{1,2}.

Entanglement has been successfully expressed in terms of correlation
functions \cite{3}, and it is related to magnetic susceptibility \cite{4}.
Thermal quantum correlations play an important role in antiferromagnetic systems that have
rich phase diagrams, and in particular in low-temperature magnetism resulting in interesting
phenomena such as magnetization plateaus.

Great interest in quantum frustrated systems has surrounded the experimental
observation of the $1/3$ magnetization plateau in the diamond chain
compound of natural material azurite $Cu_{3}(CO_{3})_{2}(OH)_{2}$, with two broad
peaks observed at around $20$ and $5 K$ in both magnetic susceptibility and
specific heat \cite{5}. High magnetic field $Cu$ NMR spectra
were used for determination of the local spin polarization in the $1/3$ magnetization plateau,
specific heat, magnetic susceptibility, magnetocaloric properties, entanglement properties
of azurite and anisotropic Heisenberg-Ising chains  by applying a variety of methods such
density functional theory, the density matrix renormalization group method, transfer matrix techniques, and the Gibbs-Bogoliubov inequality \cite{6,7,8,9,10}.

The entanglement entropy ($EE$) behavior for nearest and non-nearest neighbor sites
was investigated in the Heisenberg antiferromagnetic chain \cite{11}, with
entanglement that may exist between non-nearest neighbors at low temperatures.
In the present manuscript, logarithmic negativity ($LN$) for nearest
and non-nearest neighbors is defined in the 1D spin-$1$ antiferromagnetic
Heisenberg chain with single-ion anisotropy.
The magnetic systems of 1D spin-$1$ antiferromagnets are very attractive and magnetization plateau, magnetic susceptibility, specific heat, Neel and Haldane phases, and large single-ion
anisotropy ($D$) have been extensively studied both experimentally and theoretically \cite{12c,13c,14c,15c,16c,17c,18c}.
In this manuscript we study the properties  of a one-dimensional spin-$1$ Heisenberg antiferromagnet by considering
  the $LN$ as a function of single-ion anisotropy $D$.
  We remind that the $S=1$ Heisenberg chain provided the basis of
  the conjecture by Haldane \cite{19b} for a finite gap in antiferromagnetic
  chains with integer spin, which has to be compared to the gapless
  spectrum of half-odd-integer cases. The role of single-ion anisotropy have
  been addressed \cite{19c,20c}, the Haldane phase being in general
  detected using a non-local string order parameter \cite{19d,19e}.
  When $D$ is varied there are in general three phases.
 with the Neel phase possessing a $Z_2$ symmetry and
 the Haldane phase an incomplete $Z_2 \times Z_2$ symmetry, and with a Gaussian transition
 occurring at a critical value of $D$ \cite{19c}.

The thermal negativity
for small Heisenberg clusters of frustrated spin-$1$
anisotropic model with bilinear-biquadratic exchange,
single-ion anisotropy and magnetic field has been studied in \cite{12}.
Moreover, the single-ion anisotropy properties effect on
the magnetization plateaus, magnetic susceptibility, specific heat, and
Schottky peaks of
an octanuclear nickel phosphonate cage
$Ni_{8}(\mu_{3} - OH)_{4}(OMe)_{2}(O_{3}PR_1)_{2}(O_{2}C^tBu)_{6}(HO_{2}C^tBu)_{8}$
has been theoretically discussed \cite{13}. The results are in excellent
agreement with experimental data \cite{14, 15}.
Magnetic behavior and susceptibility of homometallic molecular
$[Ni_{3}(fum)_{2}-(\mu_{3}-OH)_{2}(H_{2}O)_{4}]_{n}(2H_{2}O)_{n}$ ferrimagnet
on a diamond chain with alternating spin values $S = 2$ and $S = 1$ were
measured at low temperatures \cite{16}. The magnetization plateaus,
specific heat, Schottky peaks, superstable points in a $2D$ mapping,
and thermal entanglement (negativity) of the spin-$1$ Ising-Heisenberg
with single-ion anisotropy on infinite diamond chain for homometallic molecular $[Ni_{3}(fum)_{2}-(\mu_{3}-OH)_{2}(H_{2}O)_{4}]_{n}(2H_{2}O)_{n}$ ferrimagnet have
been studied using the transfer-matrix approach and dynamical system theory \cite{17,18,19,20}.
The magnetic, entropy and magnetocaloric properties, and the
Gr\"uneisen parameter of the mixed spin-$(1/2,1)$ two-leg model
on an Ising-Heisenberg ladder in fully anisotropic case
with different Land\'e $g$-factors have
been examined using the transfer matrix technique \cite{21}.

The homodinuclear nickel complex (NAOC) \cite{22,22bis} serves as an experimental
realization of the spin-$1$ Heisenberg dimer \cite{23}.
The bipartite thermal entanglement (negativity)
of the quantum antiferromagnetic spin-$1$ has been discussed,
and as well the negativity plateaus as a function of the magnetic field and the uniaxial single-ion anisotropy in the ground states for quantum antiferromagnetic and ferrimagnetic coupling are also shown for NAOC. Specific heat,
susceptibility, and magnetization measurements have obtained in the spin-$1$
antiferromagnetic Heisenberg chain $Ni[C_2H_8N_2]_2Ni[CN]_4$
with uniaxial single-ion anisotropy $D$ was studied \cite{24}.
One can find the behavior of the entanglement entropy $EE$ or logarithmic
negativity $LN$ for nearest and non-nearest neighbors for
low temperatures, the external magnetic field $B$ and the
single-ion anisotropy $D$.
The points $B_c$ or $D_c$ at which entanglement disappears in the ground state
have been obtained for the antiferromagnetic Heisenberg chain $Ni[C_2H_8N_2]_2Ni[CN]_4$
and many efforts have also been devoted to the study the entanglement
entropy $EE$ and logarithmic negativity $LN$
\cite{Vidal,Wang,27,28,29,30,31,32,33,34,35,36,37,38,38a,38b,38c}.
  The logarithmic negativity $LN$ is greater than the negativity.
  For this reason, it is important to determine where logarithmic negativity
  has large, non-vanishing values for nearest and non-nearest
  neighbors in the ground and low-temperature states, and to study at what
  values of the external magnetic field ($B_c$) and single-ion anisotropy
  ($D_c$) vanish.

The presentation is organized as follows. In Section II we consider the
1D antiferromagnetic spin-$1$ Heisenberg $XXX$ model with external magnetic
field $B$ and single-ion anisotropy $D$ and we give the definition of $LN$.
The properties of entanglement states, $LN$, and the corresponding
ground-state phase diagram are discussed in Section III.
The $LN$ for nearest and non-nearest neighbors on
finite chains with sizes $N= 4 - 8$ at low temperatures for different
values of external magnetic field $B$ and single-ion anisotropy $D$ are
investigated in Section IV. Finally, concluding remarks are presented in
Section V.
\section{Model and Definitions}
The Hamiltonian for the 1D antiferromagnetic spin-$1$ Heisenberg $XXX$ model with external magnetic field $B$ and single-ion anisotropy $D$ is given as
\begin{eqnarray}
   H & = & J \sum _{i=1}^N \vec{S}_{i} \cdot \vec{S}_{i+1} +
   D \sum_{i=1}^N(S_i^{z})^2+B\sum _{i=1}^N S_i^z.  \label{ham}
\end{eqnarray}
where $J = 1$ to fix the energy scale,
$D$ represents uniaxial single-ion anisotropy,
$B$ is the controllable parameter (periodic boundary conditions are assumed).
The {\it local} spin vector $\vec{S}_i$ for each site has the components of the
spin-$1$ operators:
\begin{eqnarray}
&S^{x}&=\frac{1}{\sqrt{2}} \left(\nonumber
\begin{array}{lll}
 0 & 1 & 0 \\
 1 & 0 & 1 \\
 0 & 1 & 0
\end{array}
\right), \hskip 0.4cm
S^{y}=\frac{1}{\sqrt{2}} \left(
\begin{array}{lll}
 0 & -i & 0 \\
 i & 0 & -i \\
 0 & i & 0
\end{array}
\right), \hskip 0.4cm
S^{z}= \left(
\begin{array}{lll}
 1 & 0 &  0 \\
 0 & 0 &  0 \\
 0 & 0 & -1
\end{array}
\right).
\end{eqnarray}

At thermal equilibrium, the state of the system is determined by
the density matrix
\begin{equation}
\hat{\rho}(T)=\frac{e^{-{\frac{H}{k_BT}}}}{Z}=\sum_\alpha\frac{e^{-{\frac{E_\alpha}{k_BT}}}}{Z}|\psi_\alpha\rangle \langle \psi_\alpha|,
 \label{rho}
\end{equation}
where $E_\alpha$ are the eigenvalue of the $\alpha$-th quantum many body
eigenstate and the partition function is $Z=\sum_\alpha e^{-\beta E_\alpha}$
with $\beta = {1/k_BT}$ (from now on $k_B = 1$).

For spin-$1$ system the
degree of pairwise entanglement, measured in terms of the
negativity $Ne$, can be employed to evaluate the thermal state of
concern \cite{Vidal,Wang}.
The negativity of a state $\rho_{ab}$ is defined as
\begin{equation}
\textit{Ne}=\sum_{i}\ |\mu_{i}| , \label{negat1}
\end{equation}
where the $\mu_{i}$'s are negative eigenvalues of
$\rho_{ab}^{T_{1}}$ and $T_1$ denotes the the partial pairwise
transpose with respect to the first system, i.e., for bipartite
system in state $\rho_{ab}$ it is defined as
\begin{equation}
\langle i_{1},j_{2}|\rho_{ab}^{T_{1}}|k_{1},l_{2}\rangle \equiv\langle
k_{1},j_{2}|\rho_{ab}|i_{1},l_{2}\rangle,
\label{rhoab}
\end{equation}
for any orthonormal but fixed basis. Here $\rho_{ab}$ is a partial trace
of density matrix. Definition (\ref{negat1}) is equivalent to
\begin{equation}
\textit{Ne}=\frac{\parallel\rho_{ab}^{T_{1}}\parallel_{1}-1}{2},
\end{equation}
where $||\rho_{ab}^{T_{1}}||_{1}$ is trace norm of $\rho_{ab}^{T_{1}}$
($||*||_{1}=Tr\sqrt{*^{\dag}*}$). For unentangled states
negativity vanishes, while \textit{Ne}$>$0 gives a computable
measure of thermal entanglement.

The logarithmic negativity $LN$
is an entanglement measure defined as
\begin{eqnarray}
LN= \log_2\left(\parallel\rho_{ab}^{T_1}\parallel_{1}\right)=\log_2\left(\sum_{k}|\lambda_k|\right)
\label{ln}
\end{eqnarray}
where the $\lambda_k$'s are the eigenvalues of the partially transposed
matrix $\rho_{ab}^{T_1}$.
Since values of negativity for nearest and non-nearest neighbors are small
we consider, as entanglement measure, the logarithmic negativity $LN$.
\section{Logarithmic negativity of ground-states}
We first examine the ground state of 1D antiferromagnetic spin-$1$
Heisenberg $XXX$ models with external magnetic field $B$ and
single-ion anisotropy $D$ on the finite chain with sizes $N= 4$ and $N = 6 $.
Notice that in the following we are going to study mostly even values of $N$, even though we will comment later on the case of $N$ odd. Most of the results are obtained for finite values
  of $N$, from $N=2$ to $N=8$; however we will comment on what properties are less or not all
  affected by $N$ itself.

 Before continuing, we pause to notice that of course, $N=2$
  corresponds to the case of the Heisenberg dimer \cite{22,22bis}. However,
  notice that the Heisenberg dimer $N=2$ is studied with open boundary conditions, as natural
  for systems having two sites, while the $N=2$ limit of our results corresponds to
  the spin at the site $1$ connected
two times to the spin at the site $2$. So, to retrieve exactly the same results of Refs. \cite{22,22bis} one has to use an appropriate value of $J$.

\begin{figure}[!ht]
  	\includegraphics[width=120mm]{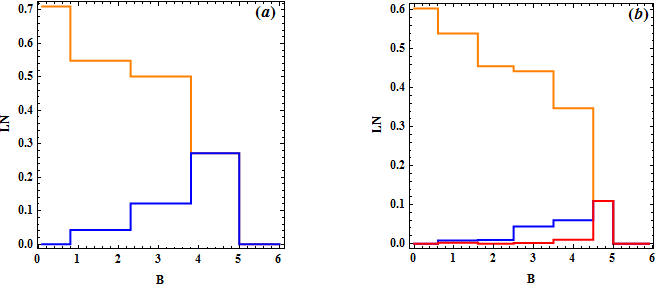}
	\caption{The logarithmic negativity $LN$ of the ground state
          as a function of external magnetic field $B$ at 
          fixed $D = 1$ (here and in the figures $J=1$).
          The upper curve shows the variation in $LN$ the amount
          of entanglement of the nearest neighbors, the middle and lowest
          curves show the same for the next-nearest and
          next-next-nearest neighbors, respectively.
          Plot (a) corresponds to the
          finite chain of size $ N = 4$, while plot (b) corresponds to
          $N = 6$.}
	\label{2dgr_b_1}
\end{figure}

The behavior of logarithmic negativity $LN$ as a function of $B$ for nearest, next-nearest and next-next-nearest neighbors is displayed in Figs.\ \ref{2dgr_b_1}, for $D = 1$.
There is a point $B_c$ in which the entanglement for $B \geq B_c$ disappears.
$B_c$ is the point of a 
quantum phase transition \cite{2}.
At the same time the critical point $B_c$  will be different for fixed values of $D$: e.g., $B_c = 3$ at $D = -1$, $B_c = 4$ at $D = 0$, $B_c = 5$ at $D = 1$
in the ground state ($T = 0$).

\begin{figure}[!ht]
  	\includegraphics[width=120mm]{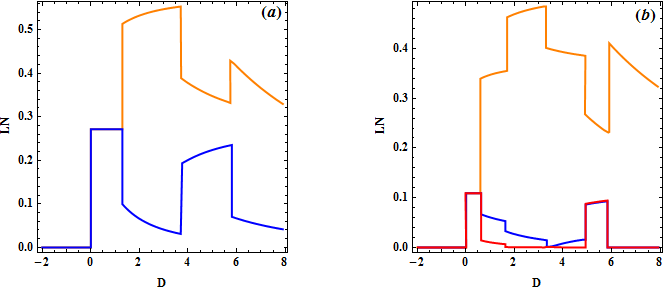}
	\caption{The logarithmic negativity $LN$ of the ground state
          as a function of the single-ion anisotropy parameter $D$
          at fixed $B = 4$. The upper curve shows the variation of $LN$
          for nearest-neighbors, the middle and lowest curves show
          the same for the next-nearest and next-next-nearest neighbors,
          respectively. Plot (a) corresponds to
          $N = 4$, and plot (b) to $N = 6$.}
	\label{2dgr_d_4}
\end{figure}

In  Fig.\ \ref{2dgr_d_4} we present the
behavior of logarithmic negativity $LN$ as a function of $D$ for nearest, next-nearest and next-next-nearest neighbors. In the ground state 
the point of a quantum phase transition from which entanglement exists
is exactly zero, $D_c = 0$, at $B = 4$ for all values of $N$.

\begin{figure}[!htbp]
  	\includegraphics[width=160mm]{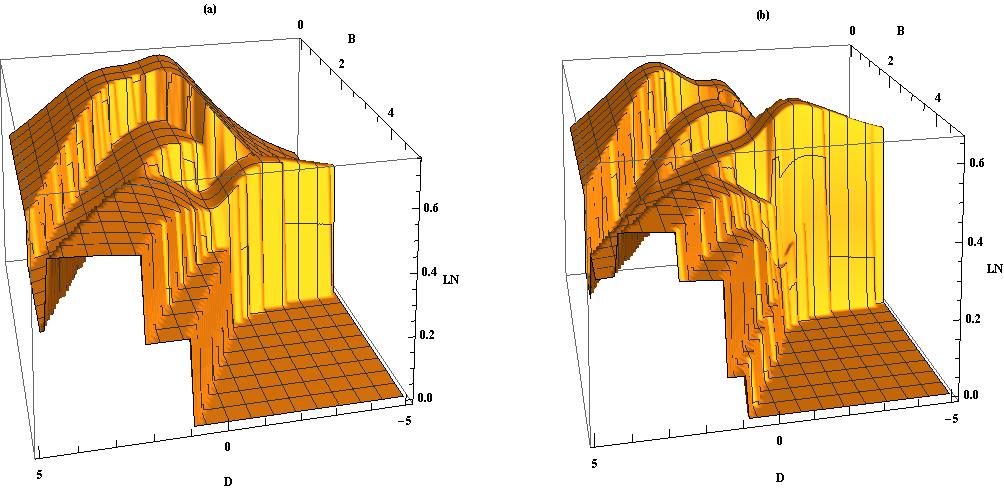}
	\caption{The logarithmic negativity  $LN$ as a function of external magnetic  field $B$ and single-ion anisotropy $D$ in ground state for finite chain of size $N = 4$ (a) and $N = 6$ (b).}
	\label{3dgr_4_6}
\end{figure}

Fig.\ \ref{3dgr_4_6} shows the plot of the logarithmic negativity $LN$ as a
function of in ground state for finite chains of sizes $N = 4$ and $N = 6$.

\begin{figure}[!htbp]
  	\includegraphics[width=160mm]{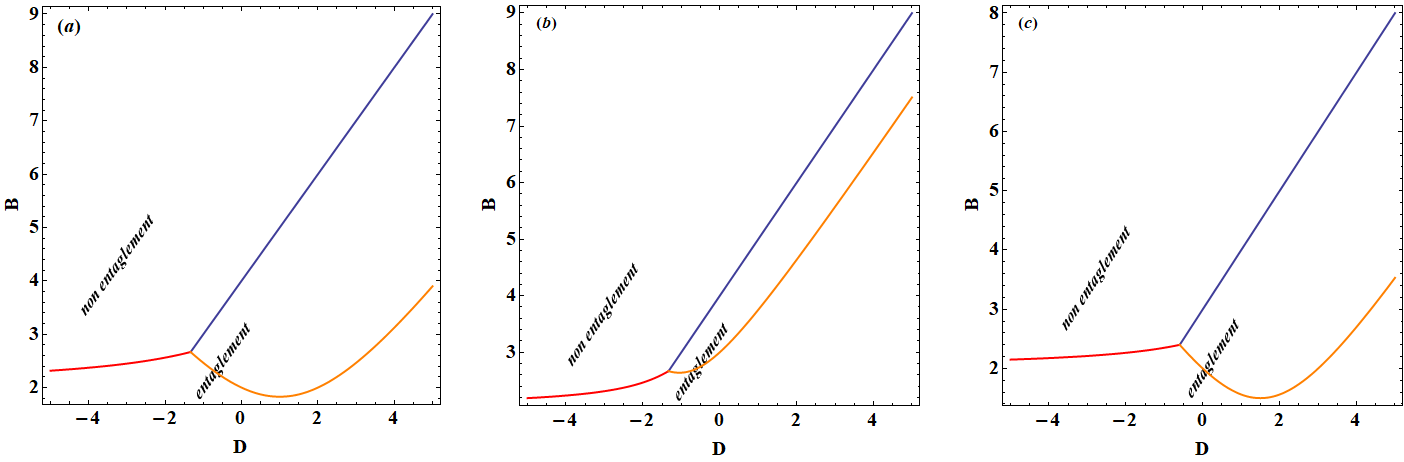}
	\caption{Ground-state phase diagram in the $B - D$ plane at
          fixed value J = 1. Plot (a) and (b) corresponds to the
          finite chain of even size $N = 2$ and $N = 4$, while plot (c) corresponds to odd
         size $N = 3$.}
	\label{ph_diag}
\end{figure}

We also found out the $T=0$ phase diagram
in the $B - D$ plane at fixed value of $J$,
see Fig.\ \ref{ph_diag} (with $J=1$).
By inspection, for $N$ even, one finds two curves separating
the two phases having vanishing and non-vanishing logarithmic negativity,
to which we will refer as the entanglement ($LN \neq 0$) and
non entanglement ($LN = 0$) phases.
The ground state of the system has a wave function corresponding
to the state with all spins down
($\Psi_{GS} = |\downarrow\downarrow....\downarrow\rangle$) in the non
entanglement region.

We have obtained the equations of these curves for $N = 2$  (see plot (a) of Fig.\ \ref{ph_diag}) :
\begin{equation}
B = 4 + D \hskip 0.5cm \left(\text{for}\,\,D > -\frac{4}{3}\right), \label{p-d1}
\end{equation}
\begin{equation}
B = \frac{1}{2}\left(3 + D + \sqrt{D^2-2 D+9}\right)  \hskip 0.5cm \left(\text{for}\,\,D < -\frac{4}{3}\right). \label{p-d2}
\end{equation}

The line (\ref{p-d1}) is found to be valid for any even $N$: this can be seen by comparing the fully polarized state $|\downarrow\downarrow....\downarrow\rangle$ and $1$-magnon state, see \cite{Yang}.

Eqs. \eqref{p-d1}-\eqref{p-d2} are in agreement with previous results for the spin-$1$ Heisenberg dimer, i.e. $N=2$, having unity spatial anisotropy \cite{22,22bis}.

While \eqref{p-d1} is valid for any even $N$, Eq. \eqref{p-d2} does depend on $N$,
so that for $D<-\frac{4}{3}$ the separatrix
between the entanglement and non entanglement phases will depend on $N$ as
well. It is important to observe that on the line (\ref{p-d1}) the energy density,
i.e., the energy of the ground state divided by $N$, is also found to
be independent on $N$. So the independence on $N$ of the
entanglement-non entanglement separatrix line is ultimately connected (and due)
to the fact that the energy density is constant.

There is a third curve, starting from the corner point, and defined for $N=2$ (plot (a) of Fig.\ \ref{ph_diag}) by
\begin{equation}
B = -1 + \sqrt{D^2-2 D+9}. \label{p-d3}
\end{equation}
The area of entanglement between this curve and the straight line (\ref{p-d3})
corresponds to the first nonzero plateau shown on the Fig.\ \ref{3dgr_4_6}, as well as the rightmost plateau
on Fig.\ \ref{2dgr_b_1}, and the leftmost one on Fig.\ \ref{2dgr_d_4}. The region
  below the orange line, on the right of the triple point ($D>-4/3$),
  corresponds to ground states that are not writable in the form \eqref{gr-wf1}, see below.

 Notice that the equation defining the orange line,
  given by (\ref{p-d3}) for $N=2$,
  generally depends on $N$. Expressions have been obtained analytically
  for $N=4$
  (not reported here): see Fig.\ \ref{ph_diag}(b),
  showing that the region between the blue and the orange line shrinks.
  but the explicit determination becomes rapidly cumbersome, already for $N=6$,
  showing the need for numerical results. However, certain features can be obtained for general
  $N$: in particular,
the 
joint intersection of the three curves, given by $D = -\frac{4}{3}$ and
$B = \frac{8}{3}$, is the same for all even values of $N$. Moreover,
  we expect that the phase diagram in plots (a) and (b) of Fig.\ \ref{ph_diag} is qualitatively valid for any finite even $N$. It would be interesting to study the dependence of the region between the blue and the orange
line as a function of $N$, for large $N$, to ascertain whether it eventually shrinks to zero.

 The plot (c) of Fig.\ \ref{ph_diag} illustrates the phase diagram for an odd value of the size of the chain: $N = 3$. We finds two curves separating the two phases having vanishing and non-vanishing logarithmic negativity:
\begin{equation}
B = 3 + D \hskip 0.5cm \left(\text{for}\,\,D > -\frac{3}{5}\right), \label{p-d4}
\end{equation}
\begin{equation}
B =\frac{1}{4} \left(\sqrt{4 D^2-12 D+25}+2 D+5\right)\hskip 0.5cm \left(\text{for}\,\,D < -\frac{3}{5}\right). \label{p-d5}
\end{equation}
The third curve, starting from the corner point, and defined for $N=3$ (plot (c) of Fig.\ \ref{ph_diag}) by
\begin{equation}
B =\frac{1}{2} \left(\sqrt{4 D^2-12 D+25}-1\right)\hskip 0.5cm \left(\text{for}\,\,D > -\frac{3}{5}\right). \label{p-d6}
\end{equation}
 The difference between equations (\ref{p-d1}) and (\ref{p-d4}) describing straight lines shows the different behavior of chains with even and odd sizes.

A very interesting behavior of logarithmic negativity occurs on the first non-zero plateau for even sizes of chain.
At values of external magnetic field $B$ and single-ion anisotropy $D$
determining this plateau, the ground state wave function is defined follow:
\begin{equation}
  \Psi_{GS} =  \frac{1}{\sqrt N}\sum_{p = 1}^{N} (-1)^{p} |\overbrace{\underbrace{\downarrow...\downarrow 0}_p \downarrow...\downarrow}^{N}\rangle,  \label{gr-wf1}
\end{equation}
where only one spin is zero ($|0\rangle$) on position $p$ and
all the other spins are down ($|\downarrow\rangle$).
The logarithmic negativity $LN$ on this plateau is determined by
only single negative eigenvalue of
$\rho_{ab}^{T_{1}}$ (\ref{rhoab}). This unique property is violated in
other regions and also on the
higher plateaus shown in Fig. \ \ref{2dgr_b_1}. At variance
the structures of the ground state wave functions are more complex for
odd $N$, although we may expect that they have the structure previously discussed
on the first non-zero plateau.

\section{Logarithmic negativity at low temperatures}
\begin{figure}[!ht]
  	\includegraphics[width=120mm]{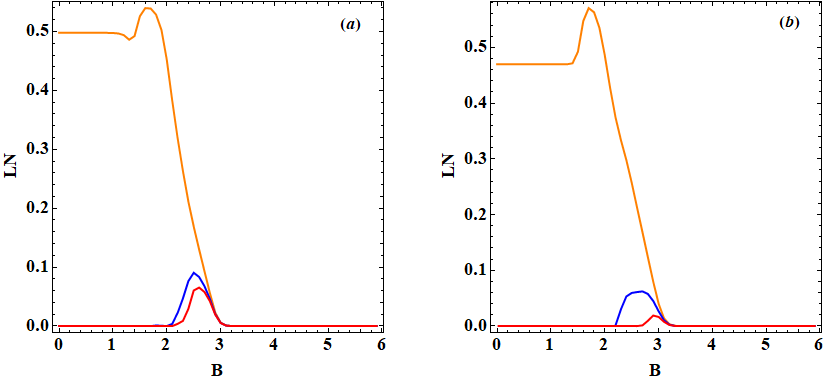}
	\caption{The logarithmic negativity $LN$ as a function of
          the external magnetic field $B$ at
          $T = 0.1$ with fixed $D = -1$. The upper curve shows
          $LN$ for  nearest neighbors, the middle and lowest curves the same for the
          next-nearest and next-next-nearest neighbors, respectively.
          Plot (a) corresponds to the finite chain of size $N = 6$, plot (b)
          to $N = 8$.}
	\label{2d_b_-1}
\end{figure}

\begin{figure}[!ht]
  	\includegraphics[width=120mm]{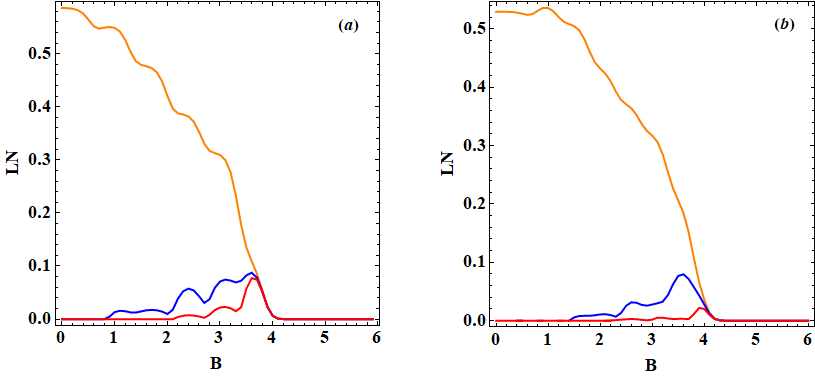}
	\caption{ The same as Fig.\ \ref{2d_b_-1}, but with fixed $D=0$.}
	\label{2d_b_0}
\end{figure}

\begin{figure}[!ht]
  	\includegraphics[width=120mm]{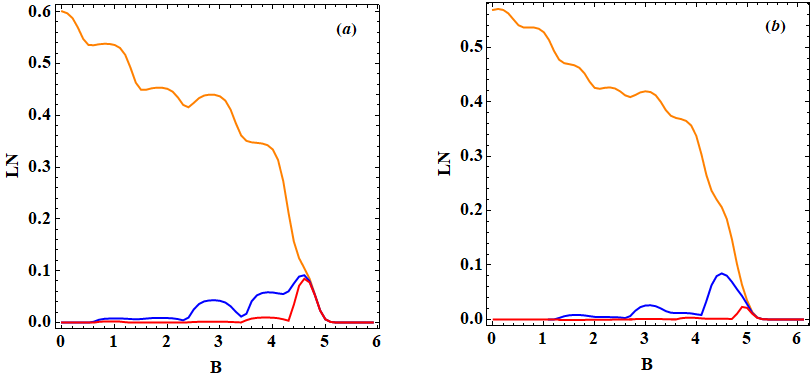}
	\caption{The same as Fig.\ \ref{2d_b_-1}, but with fixed $D=1$.}
	\label{2d_b_1}
\end{figure}

We examine in this section the 1D antiferromagnetic spin-$1$ Heisenberg $XXX$ model with external magnetic field $B$ and fixed single-ion anisotropy $D$ on the finite chain with sizes $N= 4, 5, 6, 7, 8$ at low temperatures $T$.
The behavior of logarithmic negativity $LN$ as a function of
$B$ for nearest, next-nearest and next-next-nearest neighbors
is displayed in Fig.\ \ref{2d_b_-1}, \ \ref{2d_b_0}, \ \ref{2d_b_1} at
$T = 0.1$ and $J = 1$ for $D = -1, 0 ,1$  respectively.

For finite low temperature $T$,
the entanglement disappears in a special point $B_s$.
In contrast of the ground state where the critical point $B_c$ is the
same for all values of $N$, the value $B_s$ might depend on $N$.
Moreover, $B_s$ depends on temperature and in general we find that
$B_s > B_c$.

\begin{figure}[!htbp]
  	\includegraphics[width=160mm]{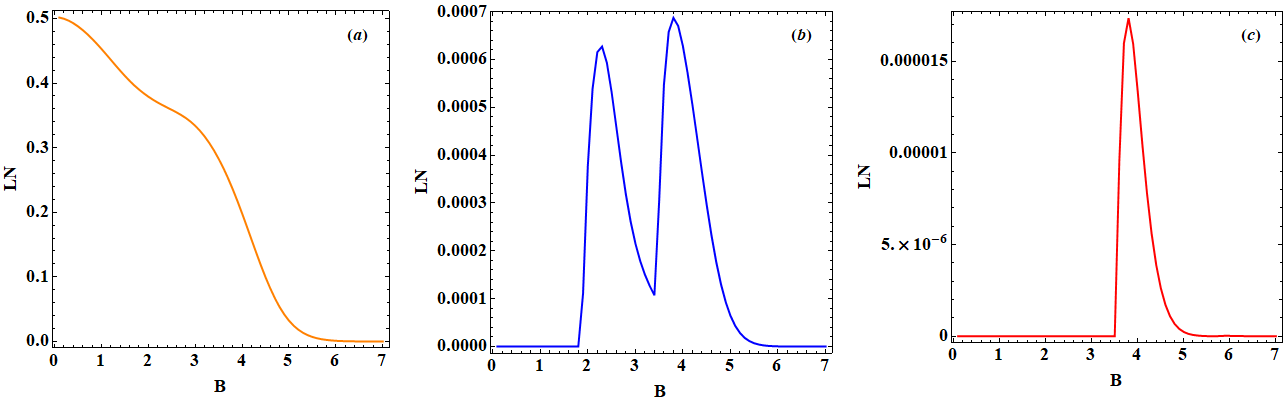}
	\caption{The logarithmic negativity $LN$ as a function of the
            external magnetic field $B$ for $ N = 6 $ at single-ion anisotropy $D = 1$ and $T = 0.5$ (with $J=1$ and $k_B=1$).
            Plot (a) shows $LN$ for nearest neighbors plot,
          while (b) and (c) corresponds to the next-nearest and
          next-next-nearest neighbors, respectively.}
	\label{2d_b_1_05}
\end{figure}

On the other hand, Figs.\ \ref{2d_b_-1}, \ \ref{2d_b_0}, and \ \ref{2d_b_1} show that non-nearest neighbors entanglement only appears after $B$
is larger than the other special point $B_e$. For the Heisenberg model
with spin-$1/2$, $B_e = B_E$, results for entanglement of next-nearest
and next-next-nearest neighbors in the ground state are universal \cite{11}.
In our case for the model with spin-$1$, the values of $B_e$ are different
for entanglement of next-nearest and next-next-nearest neighbors
(see Figs.\ \ref{2d_b_-1}, \ \ref{2d_b_0}, \ \ref{2d_b_1}).
The comparison of Fig.\ \ref{2d_b_1} (a) and Fig.\ \ref{2d_b_1_05}
  (where $D=1$ and $T=0.5$, with $J=1$ and $k_B=1$) clearly shows that the value of $B_s$ depends on temperature.

\begin{figure}[!htbp]
  	\includegraphics[width=160mm]{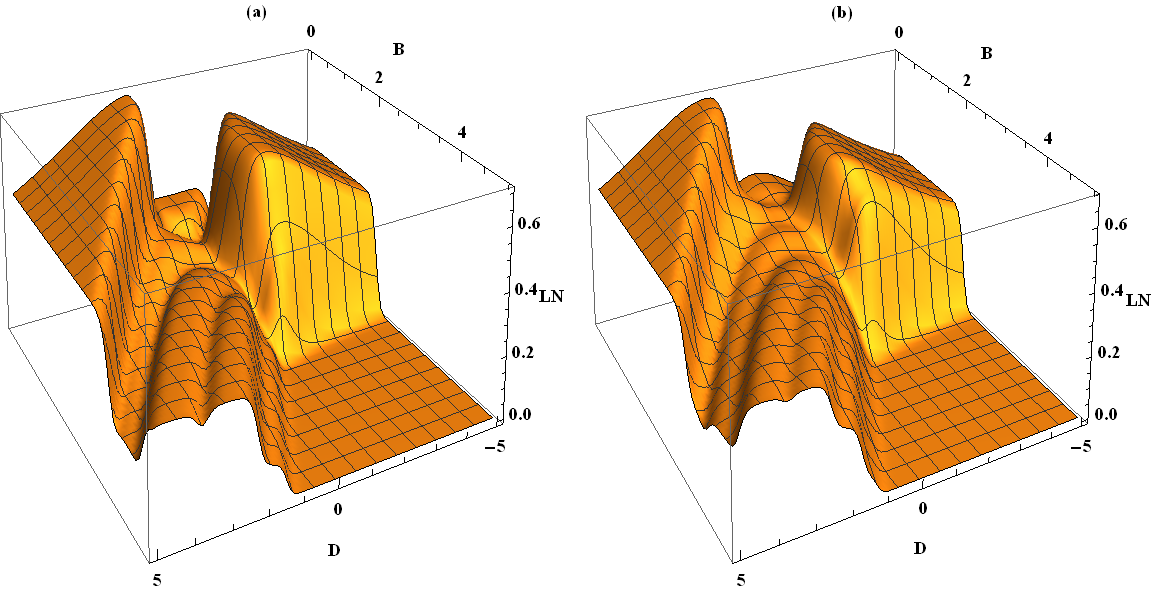}
	\caption{The logarithmic negativity $LN$ as a function of the
          external magnetic field $B$ and single-ion anisotropy $D$ at
          $T = 0.1$. Plot (a) corresponds to the finite chain of size $ N = 5$, plot (b) to $N = 7$.}
	\label{3d_5_7}
\end{figure}

\begin{figure}[!htbp]
  	\includegraphics[width=90mm]{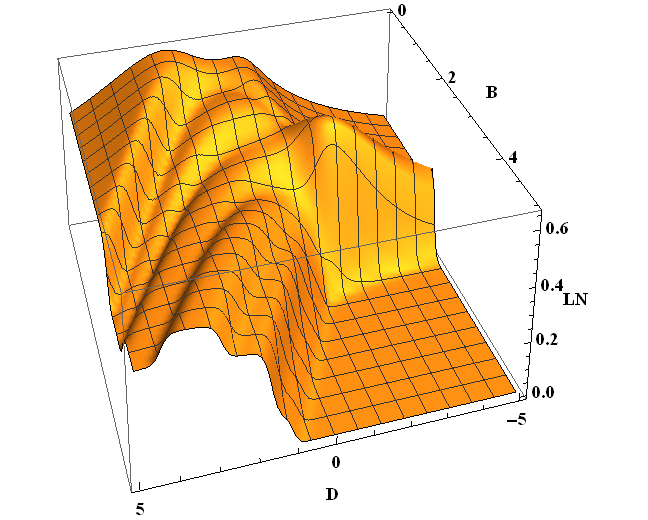}
	\caption{The logarithmic negativity $LN$ as a function of the external magnetic field $B$ and single-ion anisotropy $D$ at $T = 0.1$ for
          a finite chain of size $N = 6$.}
	\label{3d_6}
\end{figure}

As we can see from Fig.\ \ref{3d_5_7}, the behavior of logarithmic negativity
$LN$ is as expected qualitatively similar for chains of sizes $N = 5, 7$ and
different for the chain of size $N = 6$ (see Fig.\ \ref{3d_6}).
We do not provide results here for other even sizes $N = 4$ and $N = 8$,
but they are qualitatively similar to $N = 6$.

\begin{figure}[!htbp]
  	\includegraphics[width=140mm]{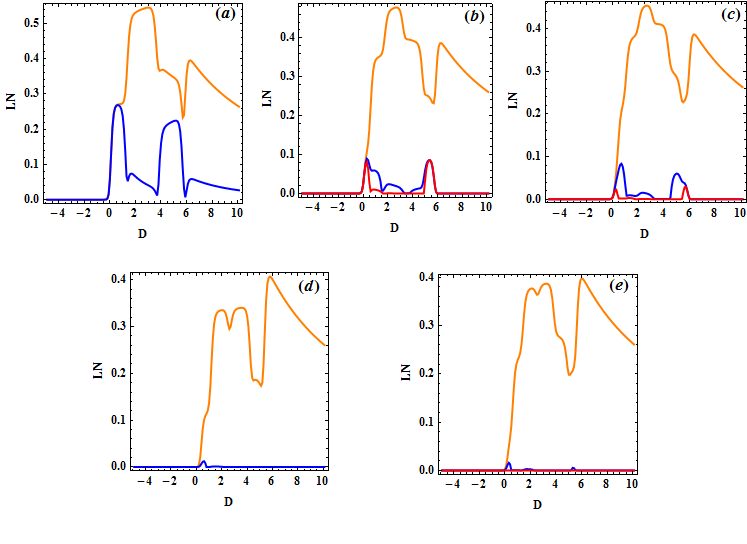}
	\caption{The logarithmic negativity $LN$ is plotted as a function of the single-ion anisotropy parameter $D$ 
          for $T = 0.1$ and $B = 4$.
          The upper curve shows $LN$
          for nearest neighbors, the middle and lowest curves show the same for the next-nearest and next-next-nearest neighbors, respectively.
          Plot (a) corresponds to the finite chain of size $ N = 4$,
          (b) to $N = 6$, (c) to $N = 8$, (d) to $N = 5$, and
          plot (e) to $N = 7$.}
	\label{2d_d_4}
\end{figure}

Now we will consider the behavior of the logarithmic negativity $LN$ of
the 1D antiferromagnetic spin-$1$ Heisenberg $XXX$ model as a function of the
single-ion anisotropy parameter $D$ with fixed external magnetic $B$.
The Fig.\ \ref{2d_d_4} shows the dependence of the entanglement $LN$ of the nearest neighbors,
next-nearest and next-next-nearest neighbors on $D$ with 
$T = 0.1$ and $B = 4$ for finite chains of sizes
$N = 4, 5, 6, 7, 8$.
Again we can see from Fig.\ \ref{2d_d_4} the behavior of
LN is qualitatively similar for chains of even sizes and
different for chains of odd sizes.

There is another special point $D_s$ starting from which entanglement exist.
The point $D_s$ is the same for the entanglement of the nearest neighbors
and for the non nearest neighbors. The value of $D_s$ is found to be
weakly dependent on $N$ (see Fig.\ \ref{2d_d_4}). We have tested this property for various values of external magnetic  fixed $B$ at low temperatures.

\begin{figure}[!htbp]
  	\includegraphics[width=160mm]{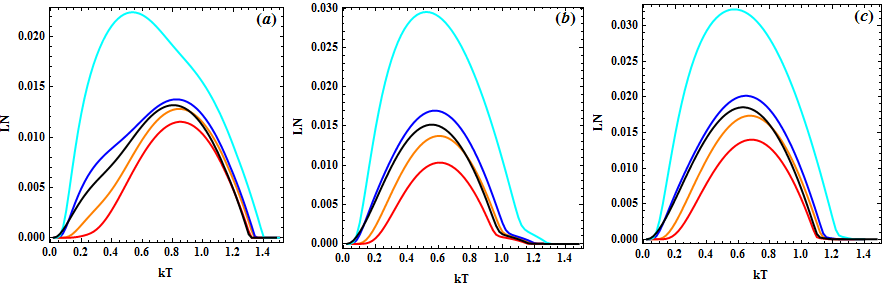}
	\caption{The nearest-neighbors logarithmic negativity $LN$ as a
function of $T$ for various values of $N$ (from top to bottom, $N = 4$, $N = 6$, $N = 8$, $N = 7$ and $N = 5$) at (a) $B = 3.2$, $D = -1$; (b) $B = 4.2$, $D = 0$; (c) $B = 5.2$, $D = 1$.}
	\label{2d_t}
\end{figure}

Let discuss now the properties of logarithmic negativity $LN$ as
function of the temperature $T$ for an external magnetic field slightly
larger than $B_s$ and at a fixed single-ion anisotropy, for the case
of nearest neighbors. In Fig.\ \ref{2d_t} we plot 
the dependence of $LN$ on $T$ for $N = 4$ $-$ $8$ and at
fixed single-ion anisotropy ($D = -1, 0, 1$).  One sees that the different
  curves tend to merge at the same value of $T$, which for the considered parameters, is
  order of $\sim 1.3$. We do not show non nearest-neighbour logarithmic negativity $LN$,
  but a similar behaviour is observed. These results can be understood from the previous finding
  that the logarithmic negativity near $B_S$ is qualitatively the same, so the
  results depicted in Fig.\ \ref{2d_t} confirm such a scenario when the field slightly
  larger than $B_s$. This is ultimately related to the fact that for
  $B$ larger than $B_s$ the system is entering a region without entanglement.
The figure  also shows
that $LN$ for even $N$ decreases, while for odd $N$ states increases with $N$.  This
  is also expected and it takes place also for spin-$1/2$, see Fig.4 of \cite{11}, and
  it can be understood by observing that for large $N$ the logarithmic negativity should tend
  to the same value irrespectively from the fact $N$ is even or odd.

\section{Conclusion}
\label{Conclusion}
The study of logarithmic negativity $LN$ behaviour for nearest and non-nearest neighbors
in quantum spin systems is a rich
and active field of research. We have investigated here the
1D antiferromagnetic spin-$1$ Heisenberg $XXX$ model with external magnetic field $B$
and single-ion anisotropy $D$. We first studied the logarithmic negativity for nearest
and non-nearest neighbors on the finite chain with sizes $N =4,6$ in the ground state.
Critical points $B_c$ and $D_c$ are also found, at which the entanglement measure disappears at zero temperature. The phase diagram is presented 
in the $B - D$ plane, where the entanglement and non-entanglement phases are separated
and a specific triple point is demonstrated.
In the ground state, plateaus of logarithmic negativity $LN$ for sizes $N = 4, 6$ are
plotted as a function of $B$ and $D$. The wave function corresponding to first non-zero
plateau for all even $N$ is found.
It would be interesting as future work to study
the ground state wave function for odd $N$.

A discussion of the behavior of logarithmic negativity $LN$ on finite chains
with sizes $N = 4 - 8$ for nearest, next-nearest and next-next-nearest neighbors
as a function of $B$ and $D$ is presented at low temperatures.
The entanglement disappears at the special point $B_s$. The latter is larger
than the critical point $B_c$ in the ground state and might depend on $N$.
We also discuss the properties of the logarithmic negativity $LN$
as a function of temperature $T$ at an external magnetic field close to $B_s$, and for a
fixed single-ion anisotropy.
The difference in behavior logarithmic negativity LN for even and odd chain sizes can be seen in Figs.\ \ref{3d_5_7}, \ref{3d_6}, \ref{2d_d_4}, which is typical for antiferromagnetic interaction
The nearest-neighbor logarithmic negativity
increases from zero temperature to $T=1.4$ and reaches a maximum, which as an example is
less than $0.035$ for $B = 3.2$, $D = -1$; $B = 4.2$, $D = 0$; $B = 5.2$, $D = 1$ for sizes $N=4,6,8,7,5$.
 For chains of even size, the logarithmic negativity LN as a function of temperature decreases with increasing size N and increases for odd sizes (see  {Fig.\ \ref{2d_t})}.

Our results are based on direct diagonalization of the Hamiltonian on finite chains for $N \le 8$.
We also have briefly mentioned experimental applications. The theoretical results obtained
from studying the behavior of logarithmic negativity $LN$ for nearest and non-nearest neighbors at low temperatures may allow, in the near future, the experimental confirmation of interesting effects
like the entanglement disappearance in the special point for the external magnetic field $B_s$
and single-ion anisotropy $D_s$ for a spin-$1$ antiferromagnetic polymer \cite{24}.

\section{Acknowledgments}
\label{Acknowledgment}
N.A. acknowledge the receipt of the grant in the frame of the research
projects No. SCS 21AG-1C006 and  No. SCS 23SC-CNR-1C006.  N.A. and A.T.
acknowledge support from the CNR/MESRA project "Statistical Physics of
Classical and Quantum Non Local Hamiltonians: Phase Diagrams and
Renormalization Group".



\end{document}